\documentclass[epj]{svjour}
\usepackage{graphicx}
\usepackage{times}
\usepackage{bm}
\usepackage{amsmath}
\usepackage{amssymb}
\DeclareMathOperator{\var}{var}
\DeclareMathOperator{\re}{Re}

\begin{document}

\title{Short-distance wavefunction statistics in 
one-dimensional Anderson localization}
\author{H. Schomerus \and M. Titov}
\institute{Max-Planck-Institut f\"ur Physik komplexer Systeme,
N\"othnitzer Str. 38, 01187 Dresden, Germany
}
\date{September 2003}

\abstract{
We investigate the short-distance statistics of the local density of states
$\nu$ in
long one dimensional disordered systems, which display Anderson
localization.  It is shown that the probability distribution function
$P(\nu)$
can be recovered from the long-distance wavefunction statistics,
if one also uses parameters that are irrelevant from the perspective of
two-parameter scaling theory.
}
\maketitle

\section{Introduction}

Wave localization in a disordered potential is the most striking hallmark
of systematic interference by multiple coherent scattering \cite{sheng,berk,Anderson,review1,review2,review3,review4,review5}.
Systematic constructive interference in a spatially localized region
results in a confinement of the wavefunction, which decays
exponentially away from the localization center (with a decay length $l_{\rm loc}$, the localization length),
in contrast to the extended waves in constant or spatially periodic potentials.
Localization comes along with
large fluctuations of the wavefunction, which can be induced
by changing the disorder configuration.
The wavefunction statistics can be probed, e.g., globally across a 
system of finite length $L_{\rm sys}$ by the
dimensionless conductance (transmission probability) $g$,
and inside a semi-infinite system ($L_{\rm sys}=\infty$)
by the local density of states $\nu$
at a distance $L_{\rm open}$ to the opening.

Theories of localization often focus on
the long-distance wavefunction statistics, where a high degree of
universality prevails. For instance, distribution functions
are restricted to log-normal forms as a consequence of the central limit
theorem, which leads to two-parameter scaling (TPS)
\cite{shapiro2}. Consequentially, for the local density of states, 
the probability distribution function $P(\nu)$ 
is characterized by the mean
logarithm $C_{1}^{(\nu)}\equiv -\langle \ln
\nu\rangle$ and its variance $C_{2}^{(\nu)}\equiv\var\ln \nu$.
The TPS observation has found many applications
\cite{markoskramer1993,Deych1,Deych2,Deych3,luan,kantelhardt,queiroz}.
An even enhanced degree of universality arises
in the random-phase approximation
(RPA), where single-parameter scaling (SPS) applies
\cite{abrahams,anderson1980,alt0,slevin}, and both parameters further are
connected, e.g.\ by $C_{2}^{(\nu)}\sim 2
C_1^{(\nu)}$ for one-dimensional systems \cite{abrikosov,lifshitz,altprig}.
(It was recognized very early that SPS breaks down for strong disorder,
see, e.g., Ref.\ \cite{kappus}.)

In this paper we point out a
connection of the long-distance statistics to the short-distance
statistics in the one-dimensional Anderson model
of localization, probed by the local density of states $\nu$.
Namely, we
find that the distribution function $P(\nu)$
for short distances reliably can be approximated with the help
of parameters that are extracted from the long-distance limit, including
parameters (besides  $C_{1}^{(\nu)}$ and
$C_{2}^{(\nu)}$) that are irrelevant, from the perspective
of TPS, for the long-distance wavefunction statistics themselves.

We start this paper by an analysis of $P(\nu)$ and $P(g)$ from
the perspective of large-deviation statistics \cite{ellis},
which goes beyond the central-limit
theorem, and identify quantities $C_n$, $n\geq 3$, which are irrelevant
for the long-distance wavefunction statistics, but will turn out to be
useful for the
short-distance wavefunction statistics.
Each quantity defines its own length scale by its asymptotic slope
$c_n=\lim_{L\to\infty}dC_n/dL$ (where $L\equiv L_{\rm sys}$ for $g$ and $L\equiv L_{\rm open}$
for $\nu$), in analogy to the relation between $C_1\sim 2L/l_{\rm loc}$ and
and the localization length $l_{\rm loc}$.
The length scales obtained from $\nu$ and $g$ coincide.
The constant offsets $d_n=\lim_{L\to\infty}C_n-L c_n$
are shown to contain information on the reflection phase, which 
allows to test the RPA.

Then we discuss that $P(\nu)$ for short distances
$L_{\rm open}\lesssim l_{\rm loc}$ can be reconstructed from
the parameters $c_n$ and $d_n$.
This is in striking contrast to $P(g)$,
for which the parameters show a transient behavior for small $L_{\rm sys}$ (where wavefunctions are not yet localized),
as was pointed out very recently in Ref.\ \cite{Deych3} (see also Ref.\ \cite{schomerus2}).
Our observations for the short-distance  statistics
lead us to conclude that the cumulants
$C_n$ are useful characteristics of localized
wavefunctions, even though they are not relevant in the
long-distance limit because of TPS.

Finally, we  analytically and numerically investigate
the parameters $c_n$ and $d_n$ in various regimes of
the one-dimensional Anderson model.

The paper is organized as follows:
In Section \ref{sec:two} we describe the general implications
of large-deviation statistics for the scaling of the
distribution functions $P(g)$ and $P(\nu)$, and identify the parameters
$c_n$ ad $d_n$ in the cumulants
$C_n$, $n\geq 3$.
In Section \ref{sec:three} we specialize to the one-dimensional Anderson model.
In order to motivate subsequent considerations, 
we first illustrate
in Section \ref{sec:threea}
the length dependence of the cumulants $C_n$ by numerical
simulations.
Then (Section \ref{sec:threeb})
we briefly review the analytical theory for the asymptotic slopes
$c_n$ \cite{schomerus2,schomerus1} and extent it
to the case of competition between onsite disorder
and offsite disorder close to the band center.
We also present the theory for the asymptotic offsets $d_n$.
In Section \ref{sec:threec} we investigate the dependence of the parameters
in various regimes of the Anderson model.
Our conclusions are given in Sec.\ \ref{sec:four}.

In order to facilitate a parallel discussion
of the statistics of $g$ and $\nu$, 
we use
the common notation $L\equiv L_{\rm sys}$ when considering $g$ and $L\equiv L_{\rm open}$
when considering $\nu$. One has to bear in mind that in the latter case,
$L_{\rm sys}=\infty$
and hence one always discusses localized wavefunctions,
while in the former case this is
true only for $L\equiv L_{\rm sys}\gg l_{\rm loc}$.

\section{Large-deviation statistics\label{sec:two}}
Large-deviation statistics often is introduced as the third and final step 
in a progressively refined analysis of the asymptotic behavior of probability
distribution functions, where the first step is the
law of large numbers 
and the second is the central-limit theorem.
In localization, the law of large numbers certifies that the
Lyapunov exponent
$\gamma=C_{1}^{(g)}/2L$ is self-averaging
in the limit $L\to\infty$ \cite{oseledec},
with asymptotic value $\lim_{L\to\infty} \gamma=l_{\rm loc}^{-1}$.
The central-limit theorem delivers a statement
about the finite-length corrections
to this asymptotic value, which are characterized by
$C_{2}^{(g)}$: The variance $\var \gamma=C_{2}^{(g)}/L^2$ decreases asymptotically
as $L^{-1}$. Presently, we find it useful not address the Lyapunov
exponents, since these are defined with help of the system length $L$,
but to rely on quantities that only involve $g$ or $\nu$,
like  $C_{1}^{(g)}$, $C_{2}^{(g)}$,  $C_{1}^{(\nu)}$, and $C_{2}^{(\nu)}$.
The law of large numbers and the central-limit theorem
predict a linear growth of these quantities with $L$.
The full picture is unfolded
in the framework of large-deviation statistics \cite{ellis}:
All cumulants 
can increase linearly with length or distance, 
\begin{subequations}
\label{cumulants}
\begin{eqnarray}
C_{n}^{(g)}&=&\langle\langle (-\ln g)^n\rangle \rangle\sim c_n^{(g)} L +  d_n^{(g)} \qquad (L\gg l_{\rm loc}),~~~~
\label{cumulantsg}
\\
C_{n}^{(\nu)}&=&\langle\langle (-\ln \nu)^n\rangle \rangle\sim c_n^{(\nu)} L +  d_n^{(\nu)} \qquad (L\gg l_{\rm loc}),~~~~
\label{cumulantsnu}
\end{eqnarray}
\end{subequations}
where the coefficients $d_n$ are the subleading corrections that can be neglected in the asymptotic limit,
but will be seen to encode information on the reflection phase that allows to test the validity of the RPA.
For the conductance $g$, the linear
scaling of the cumulants $C_{n}^{(g)}$ with $L$ and the connection of the $d_n^{(g)}$ to reflection phases
also has been found in a constructive theory by Roberts \cite{Roberts:1992}.
The parameters $c_n$
can be extracted from the averages
\begin{equation}
c^{(g)}(\xi)=-\lim_{L\to\infty} \frac{1}{L}\ln \langle g^{-\xi} \rangle =\sum_n \frac{\xi^n}{n!} c_n^{(g)}
\label{eq:c}
\end{equation}
(or equivalently for $\nu$)
as function of the continuous parameter $\xi$.
Note the exponential
dependence of the moments on $L$ due to localization, in contrast
to the power-law dependence in the critical regime
around a metal-insulator transition \cite{review4}.

This paper is centered around our numerical
observation in Sec.\ \ref{sec:three}
that Eq.\ (\ref{cumulantsnu})
holds even for short distances to the opening
$L_{\rm open}\lesssim l_{\rm loc}$, and hence can be used
in regions where the central-limit theorem does not apply.
This makes the parameters $c_n$ and $d_n$ with $n\geq 3$
observable in the distribution function $P(\nu)$, while
in the long-distance behavior only $c_1$ and
$c_2$ are relevant parameters  \cite{shapiro2}.

Presently, 
analytical results for the distribution function $P(g)$ and $P(\nu)$
for short distances are only available in the regime
of single-parameter scaling. The local density of states obeys
a strict log-normal distribution for all distances 
\cite{altprig,mesodos}, and hence complies
with our central observation.
Equation (\ref{cumulantsg}) cannot be extended to short distances,
even in the regime of single-parameter scaling \cite{abrikosov};
for studies outside this regime see,
e.g., Refs.\ \cite{Deych3,schomerus2}.

\section{One-dimensional Anderson model\label{sec:three}}
The previous section \ref{sec:two} put forward
some very general arguments from large-deviation statistics.
The relevance of the asymptotically defined parameters $c_n$ (and $d_n$), $n\geq 3$
for {\em finite-distance} wavefunction statistics, and the question whether
these parameters
indeed contain information that is independent from
what is encoded in the parameters $c_1$ and $c_2$,
only can be answered by a direct investigation.
In the following we will analyze the wavefunction statistics
in the one-dimensional Anderson model
\cite{Anderson},
given by the Schr{\"o}dinger equation
discretized on a chain (lattice constant $a\equiv 1$)
\begin{equation}
t_{l-1}\psi_{l-1}+t_{l}\psi_{l+1}=(V_l-E)\psi_l
,
\label{eq:1}
\end{equation}
where the hopping matrix elements $t_l$ and the disorder potential $V_l$
are random.
We assume box distributions with
$\langle t_l\rangle=t$, $\langle V_l\rangle=0$,
$\langle t_l t_m\rangle=t^2+\frac{1}{2} D_t\delta_{lm}$,
and $\langle V_l V_m\rangle=2 D_V\delta_{lm}$.
Without any restriction we can set $t=1$, which fixes the
energy scale in the dispersion relation $E(k)=-2\cos k$ of the
clean system ($D_t=D_V=0$).
The disorder strength will be characterized by the perturbative mean-free path
\cite{Thouless1979}
\begin{equation}
\label{eq:lpert}
l_{\rm pert}=(4-E^2)/(D_V+D_t),
\end{equation}
and the balance between onsite
and offsite disorder
will be characterized by the parameter 
\begin{equation}
\delta=(D_V-D_t)/(D_V+D_t).
\end{equation}

First we will present the results of numerical simulations to illustrate
the usefulness
of the cumulants $C_n$. Next,
in order to give a flavor for the mechanism behind the asymptotic
linear growth (\ref{cumulants}) of the cumulants 
for the specific case of wavefunction localization, 
we extent the analytical theory of Refs.\ \cite{schomerus2,schomerus1}
for the asymptotic slopes $c_n$ to the case of competition of onsite- and offsite disorder
close to the band center $E=0$, and also present the theory for the offsets $d_n$.
The extension equips us with a means to violate SPS,
which is finally compared to
other means in order to determine the mutual (in)dependence
of the parameters $c_n$ and $d_n$.

\begin{figure}[t]
\includegraphics[width=0.48\textwidth]{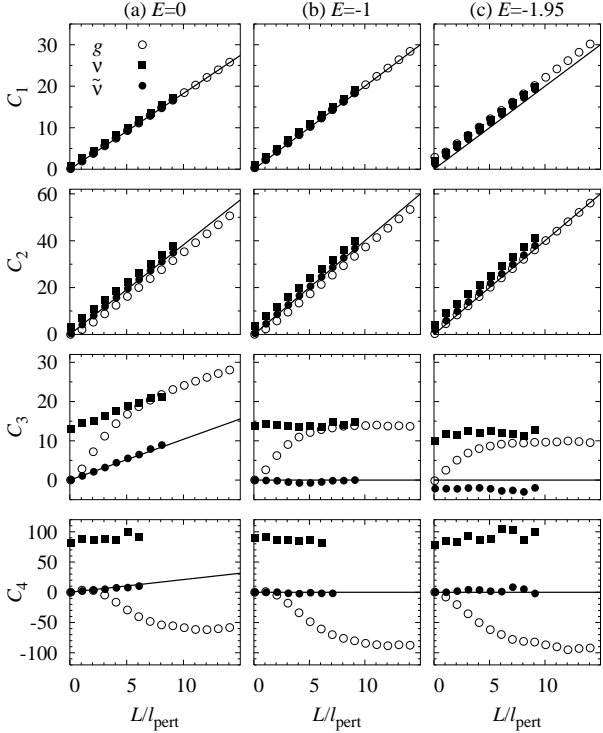}
\caption{Cumulants $C_n$
calculated from $g$ as a function of $L\equiv L_{\rm sys}$, and from $\nu$, $\tilde\nu$
as a function of $L\equiv L_{\rm open}$.
The data points are obtained by
a numerical simulation of the one-dimensional Anderson
model with onsite disorder, 
for $E=0$, $D_V=1/75$, (left panels), $E=-1$ , $D_V=1/100$ (middle panels),
and $E=-1.95$, $D_V=0.0006583$ (right panels). 
This corresponds to weak disorder, with a perturbative
mean-free path of $l_{\rm pert}=300$ in all cases
[see Eq. (\ref{eq:lpert}].
The lines are analytical weak-disorder predictions of the asymptotic linear behavior
$C_n\sim c_n L$,
taken from Ref.\ \cite{schomerus2} for $E=0$ and following the
RPA for $E=-1$ and $E=-1.95$.
}
\label{fig:1}
\end{figure}
\begin{figure}[t]
\includegraphics[width=0.48\textwidth]{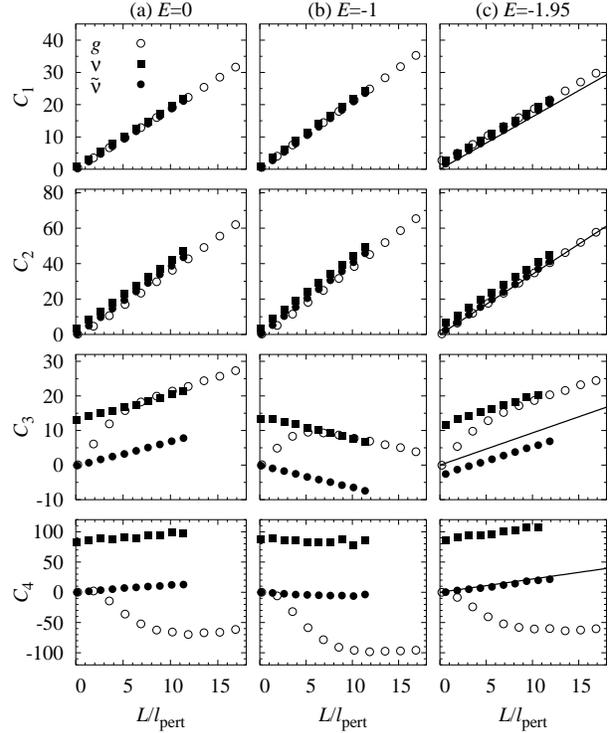}
\caption{
Same as Fig.\ \ref{fig:1}, but for stronger disorder with $l_{\rm pert}=24$:
$D_V=1/6$ (left), $D_V=1/8$ (middle), $D_V=0.00823$ (right).
The lines in the right panels (c)
are the analytic predictions for the given disorder strength close to the band edge,
taken from  Ref.\ \cite{schomerus1}.
}
\label{fig:2}
\end{figure}
\begin{figure}[t]
\includegraphics[width=0.48\textwidth]{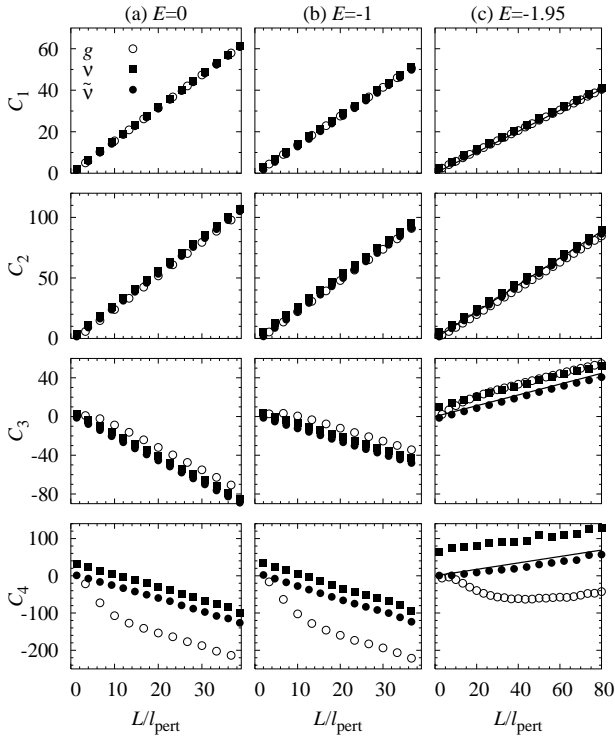}
\caption{
Same as Figs.\ \ref{fig:1}  and \ref{fig:2}, but for stronger disorder with $l_{\rm pert}=1.5$:
$D_V=8/3$ (left), $D_V=2$ (middle), $D_V=0.1316$ (right).
The lines in the right panels (c)
are the analytic predictions for the given disorder strength close to the band edge,
taken from  Ref.\ \cite{schomerus1}.
}
\label{fig:3}
\end{figure}

\subsection{Numerical illustration of the cumulants $C_n$\label{sec:threea}}

Here we illustrate
the length dependence of the cumulants $C_n$ by the
results of numerical computations
in ensembles of $10^6-10^8$ disorder realizations.
The results for 
different strengths of onsite disorder ($\delta=1$) 
are presented in Fig.\ \ref{fig:1}
($l_{\rm pert}=300$),
Fig.\ \ref{fig:2} ($l_{\rm pert}=24$), and Fig. \ref{fig:3} ($l_{\rm pert}=1.5$).
Three representative values of energy are chosen: (a) $E=0$
at the band center, (b) $E=-1$ in the SPS region, (c) $E=-1.95$ close to
the band edge.
[The constant perturbative mean free path $l_{\rm pert}$
in any figure
has been obtained by adjusting the disorder strength according to Eq.\
(\ref{eq:lpert}); for values see the figure captions.]
The significance of these three regions of energy 
will be discussed in the following Subsection \ref{sec:threeb}.
Plotted as a function of length are the cumulants $C_n$ 
calculated from
$g$ and $\nu$, as well as from the
`mesoscopic' local density of states $\tilde\nu$,
which is obtained from $\nu$ by averaging over a Fermi wavelength $\lambda_{\rm F}=2\pi/\arccos(-E/2)$
(with $\lambda_{\rm F}=4$ for $E=0$, $\lambda_{\rm F}=6$ for $E=-1$,
and $\lambda_F\approx 28$ for $E=-1.95$).
The mesoscopic density of states accounts for a limited resolution
that may be encountered in an experiment.
It discards the nodes of the wavefunction
(whose impact strongly
depends on the dimensionality of the system)
and  only captures the smoothly
varying envelope (which is more robust).

The cumulants all increase linearly 
for $L\gg l_{\rm loc}$,
and one may associate a length scale $\lim_{L\to\infty} 2L/C_n=2/c_n$
to each of them. 
The slopes  $c_n$ are identical for all 
three underlying objects, and hence
for the sets of parameters 
$\{c_n^{(g)}\}=\{c_n^{(\nu)}\}=\{c_n^{(\tilde\nu)}\}\equiv \{c_n\}$ coincide.

\begin{figure}[t]
\includegraphics[width=0.48\textwidth]{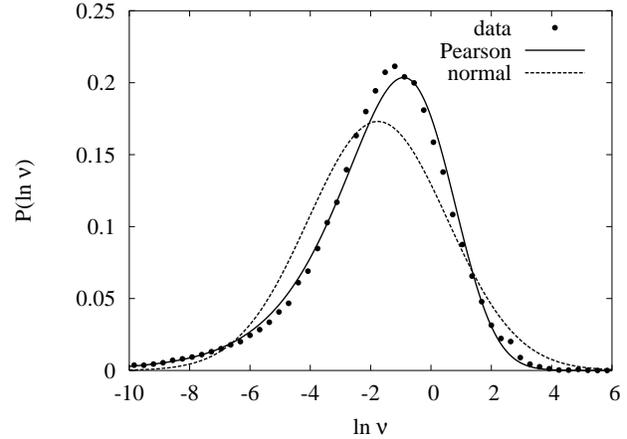}
\caption{The probability distribution function $P(\ln\nu)$ from a numerical simulation (data points)
in the Anderson model with $L=l_{\rm loc}/2$, $E=-1$, $l_{\rm pert}=24$
is compared to a normal distribution (dashed line)  with the same mean and variance as the data, and 
a generalized normal distribution from the Pearson system (solid line),
Eq.\ (\ref{eq:pearson}),
where the four free parameters are determined
from the asymptotic values of the first four cumulants.
}
\label{fig:4}
\end{figure}

As advertised above,
the cumulants $C_{n}^{(\nu)}$ and $C_{n}^{(\tilde\nu)}$
increase linearly already for small $L\lesssim l_{\rm loc}$ (moreover, the offsets for $\tilde\nu$ are vanishingly small), while
the cumulants $C_{n}^{(g)}$ become linear only after some transient
length, see Ref.\ \cite{Deych3} (these
cumulants also have a finite offset $d_{n}^{(g)}$).
This means that the asymptotically defined parameters
$c_n^{(\nu)}$ and $d_n^{(\nu)}$ can be used to estimate
the short-distance behavior of $C_n^{(\nu)}$ and $C_{n}^{(\tilde\nu)}$.
In order to estimate the distribution functions, parameters with $n\geq
3$ have to be included, since the central-limit theorem does
not yet apply for short distances.
This is displayed in Fig.\ \ref{fig:4}, which compares $P(\ln\nu)$ for $L=l_{\rm loc}/2$, $E=-1$, $l_{\rm pert}=24$
with a normal distribution, which only accounts for $C_1^{(\nu)}$ and $C_2^{(\nu)}$,
and with a generalized normal distribution (the so-called  Pearson system
\cite{pearson}),
\begin{eqnarray}
&& P(x)=C(a+bx+cx^2)^{-1/2c}
\nonumber\\&&\times \exp\left[
\frac{(b+2 c m)\arctan[(b+2cx)/\sqrt{4 ac-b^2}]}{c\sqrt{4 ac-b^2}}
\right],
\label{eq:pearson}
\end{eqnarray}
which accounts for the first four
cumulants by the four constants $a$, $b$, $c$, and $m$.
The cumulants have been reconstructed from their asymptotics
(\ref{cumulantsnu}) (hence, from the asymptotically defined quantities
$c_n^{(\nu)}$
and $d_n^{(\nu)}$), and differ from the numerical values of the data by less than three percent.

\subsection{Analytical theory for the slopes $c_n$ and offsets $d_n$\label{sec:threeb}}

\subsubsection{Slopes $c_n$}

\begin{figure}[t]
\includegraphics[width=0.48\textwidth]{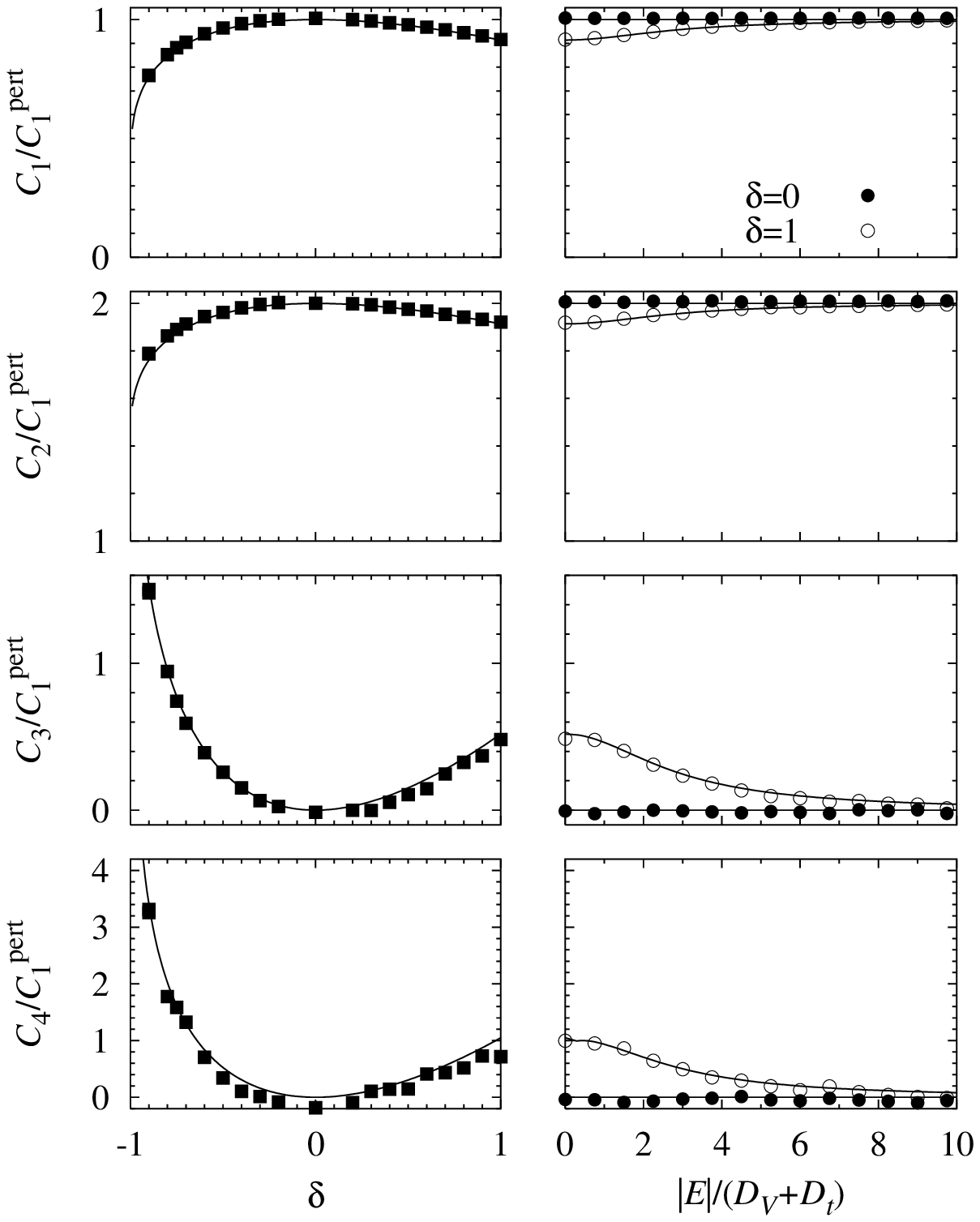}
\caption{Cumulants $C_n$ (in units of $C_1^{\rm pert}=2L/l_{\rm pert}$) in the asymptotic limit $L\gg l_{\rm loc}$,
as a function of the balance parameter $\delta$
for energy $E=0$ (left panels),
and as a function of $E$ for $\delta=0$ and $\delta=1$ (right panels).
Results of numerical simulations with $l_{\rm pert}=300$
are compared to the predictions of the analytical theory.}
\label{fig:5}
\end{figure}

Recently \cite{schomerus2,schomerus1},
we have been able to extent Halperin's phase formalism
\cite{halperin,lifshitz},
which allows to calculate $l_{\rm loc}$ and hence $c_1$,
to all slopes $c_n$. This formalism can be applied for arbitrary $\lambda_{\rm F}/l_{\rm pert}$, i.e., 
also for relatively strong disorder, as long as
$l_{\rm pert}\gg 1$ (the lattice constant, set to unity in this paper).
Other formalisms like the supersymmetric $\sigma$ model and the Berezinski{\v \i}
technique are rather more restrictive and cannot directly address the logarithm of $g$ and $\nu$.
It turned out that three different regions of energy have to be
distinguished in the one-dimensional Anderson model.
For energies $2-|E|\gtrsim D^{2/3}$ close to the band edge, corresponding to relatively strong
disorder, the RPA fails  and
the distribution function deviates from the
strict log-normal form \cite{schomerus1}.
RPA fails also
for energies  $|E|\lesssim D$ close to the band-center \cite{Stone},
and
the distribution function again deviates from the 
strict log-normal form \cite{schomerus2},
in generalization of the Kappus-Wegner anomaly of $l_{\rm loc}$
at $E=0$ \cite{kappus,derrida,goldhirsch}.
For other energies inside the band, the RPA is
justified, and SPS holds, for weak disorder.

\begin{figure*}[t]
\includegraphics[width=0.96\textwidth]{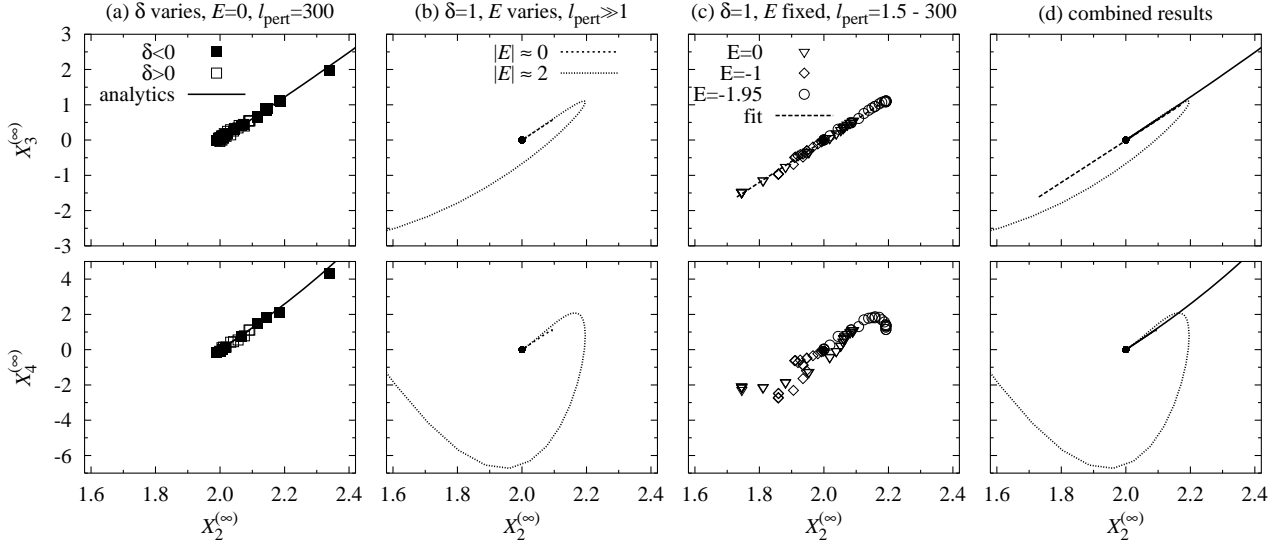}
\caption{Dependence of the parameters
$X_3^{(\infty)}$ and $X_4^{(\infty)}$ on $X_2^{(\infty)}$
for various ways to depart from SPS
[SPS conditions are indicated by the dot at coordinates (2,0)].
(a) The balance of weak onsite and offsite disorder is changed at 
the center of the band, $E=0$. The curve is the prediction of
the analytical theory,
while data points are obtained by numerical simulations.
(b) Analytical results
as energy is varied at a fixed strength of onsite disorder, with
$l_{\rm pert}\gg 1$ but $l_{\rm pert}/\lambda_{\rm F}$ arbitrary,
around the band center ($E=0$) and around the band edge ($|E|\approx 2$).
(c) Numerics and linear fits for variable strength of 
onsite disorder at three different values of the energy.
(d) Comparison of the curves in (a-c).
}
\label{fig:6}
\end{figure*}

Offsite disorder adds another means to depart from SPS at the band center,
since the balance parameter
$\delta$ interpolates between the Kappus-Wegner
anomaly at $\delta =1 $ ($D_t=0$) and the Dyson singularity at
$\delta = -1$ ($D_V=0$), which
results in total delocalization \cite{Dyson1a,Dyson1b,Dyson2a,Dyson2b,Bovier}.
In the vicinity of the band center,
we derive the following Fokker-Planck equation for the joint
distribution function  $P(u,\alpha;L)$
of $u\simeq
-\ln g\simeq -\ln \nu$,
and the phase $\alpha$ for
reflection from the system:
\begin{eqnarray}
&&(D_V+D_t)^{-1}
\frac{\partial P}{\partial L}
=
\left[
\partial_{\alpha}(\frac{\delta }{2}s_{2\alpha}-\varepsilon)
+\partial_{\alpha}^2(1+\delta c_\alpha^2)
\right.
\nonumber
\\
&& 
\left.
-\frac{1}{2}\partial_u(1+\delta
c_{2\alpha})+\frac{1}{2}\partial_u^2(1-\delta c_{2\alpha})
+\partial_u \partial_{\alpha}\delta s_{2\alpha} \right] P
,~~~~
\label{fp}
\end{eqnarray}
where $s_x=\sin x$, $c_x=\cos x$, $\varepsilon = E/(D_V+D_t)$,
and $\partial$ denotes partial derivatives.
For $\delta=1$, this equation has been used
to study the wavefunction statistics at the Kappus-Wegner anomaly
\cite{schomerus1}.
For $\delta=-1$, $E=0$, one recovers the delocalization at the Dyson anomaly.
At the balance point of onsite and offsite disorder $\delta=0$,
the variables $u$ and $\alpha$ decouple.
For $L\to\infty$
the reflection phase $\alpha$ becomes completely random, while
$P(u)$ precisely takes the form of SPS,
with asymptotic slopes $c_2=2c_1$ and
$c_n=0$ for $n\geq 3$.
Hence, somewhat surprisingly, we find that
RPA and  SPS 
hold true by a particularly simple mechanism
just in between the two 
abovementioned anomalies.

Away from the novel SPS point $\delta=0$, but for disorder still small,
the asymptotic behavior of the distribution function can be 
analyzed by introducing into Eq.\ (\ref{fp}) the large-deviation ansatz
\begin{equation}
P(u,\alpha;L)=\int_{-i\infty}^{+i\infty}\frac{d\xi}{2\pi i}
\sum_{k=0}^\infty \exp[c (\xi) L-\xi u]f(\alpha;\xi).
\label{eq:ansatz}
\end{equation} 
Here $c(\xi)=\sum_{n}\xi^n c_n / n!$ 
is the generating function of the slopes of the cumulants,
see Eqs.\ (\ref{cumulants}) and (\ref{eq:c}), and $f(\alpha;\xi)$ has to be periodic and
normalizable with respect to $\alpha$.
We arrive at a differential equation 
\begin{eqnarray}
&&\frac{c(\xi)f(\alpha;\xi)}{D_V+D_t}
=
\left[
\partial_{\alpha}(\frac{\delta}{2} s_{2\alpha}-\varepsilon)
+\partial_{\alpha}^2(1+\delta c_\alpha^2)
\right.
\nonumber
\\
&& 
\left.
+\frac{\xi}{2}(1+\delta
c_{2\alpha}-2\partial_{\alpha}\delta s_{2\alpha})
+\frac{\xi^2}{2}(1-\delta c_{2\alpha})
\right] f(\alpha;\xi)
,~~~~
\label{dgle}
\end{eqnarray}
in which the slope-generating function $c(\xi)$ appears as an eigenvalue,
while $f(\alpha;\xi)$ appears as an eigenfunction.
The slopes $c_n$ now
can be calculated iteratively by expanding $c(\xi)$ and $f(\alpha;\xi)$
order by order in $\xi$, following
Refs.\ \cite{schomerus2,schomerus1}.
Away from the SPS point $\delta=0$ but for $|E| \lesssim D_V+D_t$, the slopes $c_n$ take finite values, in compliance with
Eq.\ (\ref{cumulants}). 
Our analytical results are confirmed by numerical computations in Fig.\ \ref{fig:5}.

\subsubsection{Constant offsets $d_n$}

The offsets
$d_n^{(\nu)}\approx C_{n}^{(\nu)}(L=0)$ can be calculated by expressing the
local density of states $\nu(L)$ in terms of the reflection coefficients $r_{\rm R}$ ($r_{\rm L}$) from the 
segment of the wire to the right (left)
of the point $L$ at which $\nu$ is calculated \cite{mesodos},
\begin{equation}
\nu(L)= 
\re\frac{(1+r_{\rm L})(1+r_{\rm R})}{1-r_{\rm L}r_{\rm R}},
\end{equation}
where we normalized $\langle \nu(L)\rangle=1$ (which amounts to multiplication by a constant factor $\pi\sqrt{4-E^2}$).
For $L=0$ and the opening of the wire oriented to the left, $r_{\rm L}=0$ because there
is no reflection from the opening, and $r_{\rm R}=\exp(i\alpha)$,
where $\alpha$ is the phase of reflection from the semi-infinite system.
Hence, the numbers
\begin{equation}
d_n^{(\nu)}\approx C_{n}^{(\nu)}(L=0) =\langle\langle[
-\ln(1+ \cos\alpha)]^n\rangle\rangle
\end{equation}
characterize the distribution of the reflection phase  $\alpha$
of the semi-infinite system \cite{mesodos}, and
allow to assess the validity of the RPA,
which predicts $d_1^{(\nu)}=\ln 2 $, $d_2^{(\nu)}=\pi^2/3$, 
$d_3^{(\nu)}=12 \zeta(3)$ (with the Riemann $\zeta$ function), $d_4^{(\nu)}=14 \pi^4/15$.

The offsets  $d_n^{(\tilde\nu)}\approx C_{n}^{(\tilde\nu)}(L=0)$  vanish independently of the RPA
since in terms of the reflection matrices introduced above
\begin{equation}
\tilde\nu(L)=
\re\frac{1+r_{\rm L}r_{\rm R}}{1-r_{\rm L}r_{\rm R}},
\end{equation}
and hence $\tilde\nu(0)=1 
$.

The offsets $d_n^{(g)}$ are obtained by considering the composition law
$t_{\rm R+L}=t_{\rm R}(1-r_{\rm L}r_{\rm R})^{-1}t_{\rm L}$
for the series transmission through
two long segments R and L. The reflection coefficients now are equivalent phase factors $r_{\rm R,L}=\exp(i\alpha_{\rm R,L})$.
We equate the cumulants of both sides and insert the asymptotics (\ref{cumulants}).
The constant offsets follow as $d_n^{(g)}=(-1)^n\langle\langle \{\ln [2-2\cos(\alpha_{\rm R}+\alpha_{\rm L})]\}^n\rangle\rangle$.
In the RPA, 
$d_1^{(g)}=0$ and $d_n^{(g)}=(-1)^n d_n^{(\nu)}$. This
is clearly displayed in Fig.\ \ref{fig:1}.
Beyond the RPA, the $d_n^{(g)}$ and $d_n^{(\nu)}$ contain equivalent information on the reflection-phase distribution
function $P(\alpha)$,
but no longer are simply related.

\subsection{Independence of the parameters\label{sec:threec}}

Now we turn to the question of the mutual independence of the parameters $c_n$, 
as we violate the conditions for SPS.

A convenient set of parameters
beyond the SPS quantity $C_1$
is formed by the ratios $X_n=C_n/C_1$, which
asymptotically acquire the constant values 
\begin{equation}
\lim_{L\to\infty} X_n=X_n^{(\infty)} = c_n/c_1.
\label{x}
\end{equation}
In RPA and SPS,
only one effective parameter $C_1$ survives since 
$c_2^{(g)}=c_2^{(\nu)}=2c_1^{(g)}=2c_1^{(\nu)}$, i.e., $X_2^{(\infty)}=2$,
and moreover $c_n^{(g)}=c_n^{(\nu)}=0$ for $n\geq 3$,
which gives a picture consistent with SPS.
However, beyond this approximation the cumulants $C_n$
with $n\geq 3$ generally may increase linearly with $L$,
and hence can be of the same order as $C_1$ and $C_2$, such that all $X_n^{(\infty)}$ are of order unity.
Notice that the asymptotic value $X_n^{(\infty)}$
is well approximated by  $X_n^{(\tilde\nu)}$
even for
$L\lesssim l_{\rm loc}$, since the cumulants $C_n^{(\tilde\nu)}$
are linear already for small $L$ and the offsets $d_n^{(\tilde\nu)}$ vanish.

In Fig.\ \ref{fig:6} we
plot the asymptotic ratios of cumulants $X_3^{(\infty)}$ and 
$X_4^{(\infty)}$ as
function of $X_2^{(\infty)}$, while we vary:

(a) the balance parameter
$\delta$ at $E=0$ (a i) from 0 to 1 and (a ii) from 0 to $-1$;

(b) 
energy for fixed onsite disorder (b i) around $E=0$ and (b ii) around $|E|=2$;
and

(c) 
the disorder strength from $l_{\rm pert}=300$ to $l_{\rm pert}=1.5$
at $\delta=1$ for  the three values of energy (c i) $E=0$, (c ii) $E=-1$,
and (c iii) $E=-1.95$.

In the cases (a) and (b) we show the results of the analytical
procedure described above, while for (c) we show the result
of the numerical simulations.
For illustration of the predictive power of the theory
presented in the previous Sec.\ \ref{sec:threeb},
numerical results are also displayed for case (a).

Of particular interest is the curve for case (b ii),
for energies close to the band edge, which also applies to 
strong disorder, $D^{2/3}\gtrsim 2-|E|$ \cite{schomerus1}.
[See also the data points for case (c iii).]
In this case the curves $X_n^{(\infty)}(X_2^{(\infty)})$
depart from the seemingly unique functional behavior obtained in the other
cases.  Hence, we are led to conclude that at least for sufficiently strong
disorder $X_3^{(\infty)}$ and $X_4^{(\infty)}$ are not
uniquely determined by $X_2^{(\infty)}$. Since the SPS quantity
$C_1$ is always an independent scaling parameter, altogether more than two
quantities are needed to characterize the  distribution function
$P(\nu)$ for short distances $L\lesssim l_{\rm loc}$
(where the central-limit theorem, and hence TPS, not yet applies).

\section{Conclusions\label{sec:four}}
We observed that the {\em short-distance} statistics of localized wavefunctions
inside a long one-dimensional disordered system
can be recovered from the long-distance statistics, 
but
in general 
are characterized by more than the two parameters (a mean $C_1$ and a
variance $C_2$)
that suffice to describe
the long-distance statistics themselves.
These additional parameters have been
obtained from the higher cumulants $C_n$
of $\ln \nu$,
where $\nu$ is the local density in a semi-infinite system.
The additional parameters in the $C_n$
can be neglected when considering the case of weak disorder and generic
energies within the band: Then $C_1\sim C_2/2\propto L$,
and also $C_n=O(L^0)$ for $n\geq 3$ take universal values,
which results in a picture
consistent with single-parameter scaling even
in the short-distance wavefunction statistics.

With three-dimensional systems in mind,
it would be desirable to investigate the relation
of the parameters from large deviation statistics to the 
scaling parameters at the metal-insulator transition,
which may be established by multi-fractal analysis 
when this transition is approached from the localized regime.

Another potential
application of the higher cumulants
is to use them for detecting spatial correlations in the disorder,
since the higher cumulants are sensitive
to higher-point wavefunction correlations.
This offers a natural extension of a previous
investigation \cite{Brenner:Fishman:1992}, which demonstrated that
deviations from randomness
due to spatial three-point correlations (such as displayed by a 
folded Fibonacci sequence) cannot be detected by the conventional
wavefunction statistics.

\end{document}